\documentclass{jfm_modified_2021}
\usepackage{graphicx,import}
\usepackage{amsmath, amssymb}
\usepackage[inactive,tightpage]{preview}

\usepackage{microtype}
\usepackage[normalem]{ulem}
\usepackage{mathtools,bm}
\usepackage{cases}

\usepackage{esvect}

\usepackage{enumitem}

\pdfmapfile{+rsfso.map}
\DeclareSymbolFont{rsfso}{U}{rsfso}{m}{n}
\DeclareSymbolFontAlphabet{\mathscr}{rsfso}

\usepackage{pdflscape}
\usepackage{afterpage}
\usepackage{flafter}
\usepackage{rotating}
\usepackage{fancyhdr}
\fancypagestyle{lscape}{%
\fancyhf{} 
\fancyfoot[LE]{}
\fancyfoot[LO] {}
 
}

\usepackage{bm}
\usepackage{microtype}

\usepackage{hyperref}
\usepackage[usenames, dvipsnames]{xcolor}
\hypersetup{colorlinks=true,linkcolor=MidnightBlue,citecolor=MidnightBlue}
\usepackage{natbib}
\usepackage{array}
\usepackage{booktabs}
\usepackage{multirow}
\usepackage{siunitx}
\usepackage{mathtools}

\usepackage{cleveref}
\usepackage{caption}
\usepackage{subcaption}
\usepackage{tabularx}
\usepackage{makecell}
\newcolumntype{Y}{>{\centering\arraybackslash}X}

\usepackage{tikz}
\usepackage{pgfplots}
\usetikzlibrary{backgrounds, calc, patterns.meta, shadows, arrows, patterns}
\usepackage[outline]{contour} 
\contourlength{0.15em}

\usepackage{lscape}

\newcommand*{\e}{\mathrm{e}}
\newcommand*{\Oh}{\mathcal{O}}

\newcommand{\dt}[1]{\frac{\mathrm{d} #1}{\mathrm{d} t}}
\newcommand{\dtp}[1]{\frac{\mathrm{d} #1}{\mathrm{d} t'}}
\newcommand{\dd}[2]{\frac{\mathrm{d} #1}{\mathrm{d} #2}}

\newcommand{\pdx}[1]{\frac{\partial #1}{\partial x}}
\newcommand{\pdt}[1]{\frac{\partial #1}{\partial t}}

\shorttitle{Asymptotic analysis of PDM}
\shortauthor{Morawiecki and Trinh}

\title{Asymptotic differences between a lumped probability-distributed rainfall-runoff model and a physical benchmark model}

\author{Piotr Morawiecki\corresp{\email{pwm27@bath.ac.uk}}
 \and Philippe Trinh\corresp{\email{hppt20@bath.ac.uk}}}

\affiliation{
Department of Mathematical Sciences, University of Bath, Bath BA2 7AY, UK
}

\pubyear{XXXX}
\volume{YYY}
\pagerange{xxx--xxx}
\date{\today~[Draft]}

\newcommand{\tsat}[0]{t_\mathrm{crit}}
\newcommand{\qsat}[0]{q_\mathrm{crit}}

\begin{document}

\maketitle

\begin{abstract}
Typically, physical and conceptual rainfall-runoff models are developed independently and are based on different, though not entirely incompatible, governing principles. In this work, we perform a systematic asymptotic analysis of a typical conceptual rainfall-runoff model and compare its results to a physical model. The setting for this experiment is a geometrically simplified benchmark catchment consisting of a two-dimensional hillslope. For demonstration purposes, we analyse the lumped Probability Distributed Model (PDM) formulated by~\cite{lamb1999calibration}.
Our analysis reveals fundamental differences between the PDM and physical models, both in qualitative and quantitative behaviours, particularly when studying overland-flow-dominated catchments. For example, our study highlights the fact that the PDM model overestimates the rate of river flow growth during highly intensive rainfalls. Additionally, we observe that the PDM model cannot accurately predict peak flows in different seasons, characterised by a wide range of mean precipitation/evapotranspiration rates.
We argue that this analytical approach of identifying fundamental differences between models can serve to understand model limitations and develop more theoretically-justified rainfall-runoff models.
\end{abstract}

\section{Introduction}

\noindent Traditionally, the models used for predicting river flows can be divided into three main classes: physical, conceptual, and statistical models~\citep{sitterson2018overview}. Physical models are based on equations of hydrodynamics to describe surface and subsurface flow. However, they have high data and computing power requirements, which is why conceptual rainfall-runoff models are often preferred; these latter models take advantage of physically intuitive concepts such as the separation of surface and subsurface flows, but represent physics with simpler, often less theoretically justified, mathematical formulations.

The goal of this paper is to investigate how accurate are predictions of such conceptual model, as compared to hydrographs produced by a physical benchmark model. We focus on the Probability Distributed Model (PDM) by \cite{moore1985probability}, and in particular its implementation in a six-parameter lumped model by \cite{lamb1999calibration}. Our choice is motivated by the simplicity of this class of conceptual models, as well as its relative importance in hydrological modelling in the UK. PDM is used as a subcomponent of other rainfall-runoff models; for example, a lumped PDM is used in the UK Environmental Agency National Flood Forecasting System \citep{moore2007pdm}. Similar, PDM is used within a distributed Grid-to-Grid model by \cite{bell2007development}, used for the operational flood forecasting services of England, Wales, and Scotland \citep{cole2010grid, price2012operational}, by the Australian Bureau of Meteorology \citep{acharya2019evaluation, wells2019distributed},
and within a wide range of research projects, \emph{e.g.} by \cite{bell2009use}, \cite{bell2018marius}, \cite{formetta2018estimating}, and \cite{kay2021climate}.

The PDM we use in this work uses a probability distributed model for the storage capacity, as introduced by \cite{moore1985probability}. Instead of assuming a fixed soil capacity, this approach represents the soil capacity using a probability distribution. Consequently, during the rainfall, the PDM models the feature where parts of the catchment can become saturated earlier than others, contributing to the generation of a fast overland flow. We shall introduce this model more formally in \cref{sec:PDM}.

\cite{lamb1999calibration} demonstrated that, after a proper calibration, a simple six-parameter PDM can accurately predict discharge for Frome and Ythan catchments in the UK with up to a 10-year return period. Note that though this demonstrates that the model can be successfully applied, the conclusions are restricted to the data investigated. The conclusions of such numerical studies are hard to be generalise, and an open question remains about the limits of applicability of investigated models---\emph{i.e.} are there conditions under which predictions become inaccurate? There is significant discussion of model uncertainty within the literature by \emph{e.g.} \citet{klemevs1986operational}, \citet{beven2018hypothesis} and \citet{beven2019make}.

In this paper, we follow an alternative approach, which seeks to compare predictions obtained by different models when studied in the specific context of a simple synthetic benchmark scenario. Such approach was used, for example, in the two-part intercomparison study by \cite{maxwell2014surface} and \cite{kollet2017integrated}; in these works, predictions of different physical computational models were compared over several simple geometries, including a single hillslope or a V-shaped catchment. Here we use the physical benchmark model developed in an earlier three-part study \citep{paper1,paper2,paper3}, where analytic asymptotic solutions are possible. We briefly summarise this model in \cref{sec:physical_benchmark}.

\begin{figure}
    \centering
    \begin{tikzpicture}[font=\small]

\definecolor{color1}{RGB}{67,162,202}

\tikzset{legendstyle/.style={anchor=west,inner sep=0pt, outer sep=0pt,text width=180, scale=0.85}}

\begin{axis}[
    width=0.35\linewidth,
    height=0.35\linewidth,
    xmin={0},
    xmax={24},
    ymin={0},
    ymax={1.5e-4},
    xtick={0,6,12,18,24},
    xlabel={Time $t$ [h]},
    ylabel={Total flow $Q$ $\left[\mathrm{m/s^2}\right]$},
    title={Low saturation (20\%)},
    at={(0,0)},
    ]

\addplot [line width=1.5pt, color1] table [mark=none, x index=0, y index=5, col sep=comma] {DATA/benchmark_r0_dependence.dat};

\addplot [line width=2pt, color1, dashed] table [mark=none, x index=0, y index=5, col sep=comma] {DATA/pdm_r0_dependence.dat};

\end{axis}

\begin{axis}[
    width=0.35\linewidth,
    height=0.35\linewidth,
    xmin={0},
    xmax={24},
    ymin={0},
    ymax={1.5e-4},
    xtick = {0,6,12,18,24},
    yticklabels={,,},
    title={Medium saturation (50\%)},
    at={(0.3\linewidth,0)},
    ]

\addplot [line width=1.5pt, color1] table [mark=none, x index=0, y index=3, col sep=comma] {DATA/benchmark_r0_dependence.dat};

\addplot [line width=2pt, color1, dashed] table [mark=none, x index=0, y index=3, col sep=comma] {DATA/pdm_r0_dependence.dat};

\end{axis}

\begin{axis}[
    width=0.35\linewidth,
    height=0.35\linewidth,
    xmin={0},
    xmax={24},
    ymin={0},
    ymax={1.5e-4},
    xtick = {0,6,12,18,24},
    yticklabels={,,},
    title={High saturation (80\%)},
    at={(0.6\linewidth,0)},
    ]

\addplot [line width=1.5pt, color1] table [mark=none, x index=0, y index=1, col sep=comma] {DATA/benchmark_r0_dependence.dat};

\addplot [line width=2pt, color1, dashed] table [mark=none, x index=0, y index=1, col sep=comma] {DATA/pdm_r0_dependence.dat};

\end{axis}

\end{tikzpicture}
    \caption{The comparison of the physical benchmark hydrographs (solid lines) and the hydrographs predicted by the PDM (dashed lines). Each graph represents a different fraction of the soil being initially saturated. Sec. \ref{sec:P0_dependence} includes more information about the settings used to generate this graph.}
    \label{fig:r0_dependence_intro}
\end{figure}
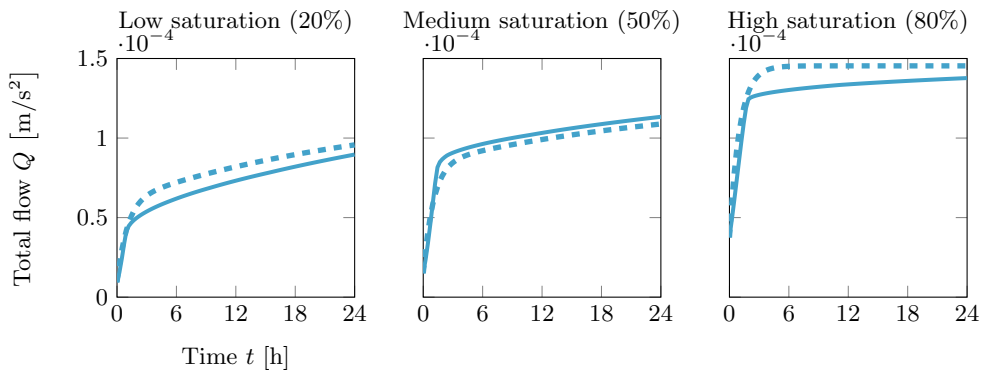

As we shall demonstrate in this paper, the PDM implementation by \cite{lamb1999calibration} can be calibrated to reconstruct hydrographs generated with the physical benchmark model; however in doing so, we also observe some significant qualitative and quantitative differences the two models. A particularly interesting conclusion we shall draw is that without model calibration, PDM cannot properly reconstruct the shape of the hydrographs in seasons that are characterised by different mean precipitation values (and consequently soil saturation levels). Consequently, the PDM tends to either overestimate the river flow during the most humid and most dry seasons; or the model tends to underestimate flow for intermediate temporal values. A comparison is shown in \cref{fig:r0_dependence_intro}.

These and other quantitative and quantitative differences are discussed in-depth in further parts of this paper; a succinct summary of key similarities and differences can be found in \cref{tab:summary} of \cref{sec:conclusions}.

\section{Physical benchmark scenario}
\label{sec:physical_benchmark}



\noindent The physical benchmark scenario we shall used to compare with the results of the PDM model is based on the coupled surface-subsurface model formulated in \citep{paper2}, which under the shallow-depth regime, possesses analytical approximations in the case of overland-dominated catchments, as presented in \citep{paper3}. In this section, we provide a review of the physical benchmark model presented in these two works; further details can be found in \citet{piotr_thesis}. 

\begin{figure}
    \centering
    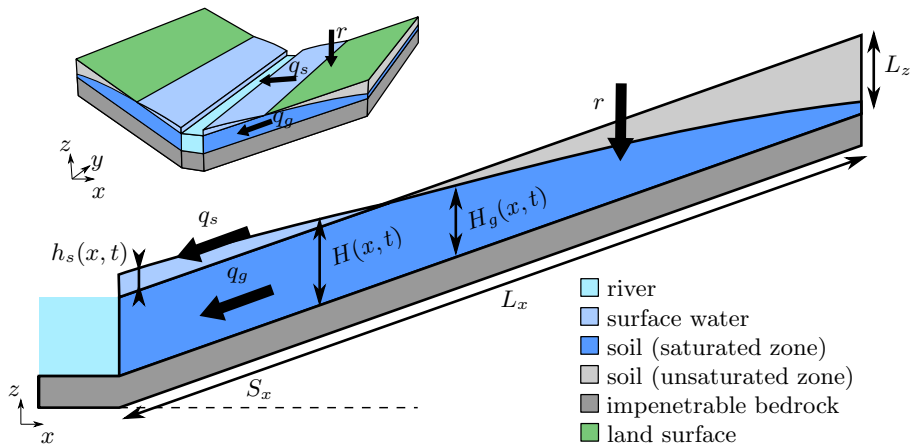
    \caption{Simplified scenario representing overland-dominated catchments.}
    \label{fig:1D_hillslope}
\end{figure}

We consider a model of a simplified catchment with hillslopes consisting of a thin layer of soil constrained by impenetrable bedrock, as shown in \cref{fig:1D_hillslope}. For simplicity, we assume that all physical parameters such as soil depth, surface slope, and hydraulic conductivity are constant along the hillslope. Initially, the system begins in an initial state that corresponds to the steady-state flow configuration for a hillslope exposed to mean constant rainfall of $r_0$. Then at time $t > 0$, a constant intensive rainfall $r > r_0$ is applied to the hillslope. The numerical values of physical parameters are chosen to represent the behaviour of low-productive catchments in the UK (cf. \citealt{paper1}); the numerical values of parameters for the model presented below can be found in Table 1 of \cite{paper3}.

The mathematical model uses Richards equations for the subsurface flow and the Saint Venant equations for the surface flow. If the aquifer is considered to be thin (i.e. $L_z\ll L_x$ in \cref{fig:1D_hillslope}), the subsurface flow can be considered under a shallow-water approximation, leading to the Boussinesq equation. We let $H(x, t) = H_g(x, t) + h_s(x, t)$ denote the sum of the subsurface and surface heights. As shown in \cite{paper3}, the evolution of the total height, $H$, is given by:
\begin{equation}
    \label{eq:dHdt_dimensional}
    \pdt{H} =
    \begin{cases}
        f(x)^{-1}\left[\pdx{}\left(K_s H\pdx{H} + K_s H S_x\right) + r\right] & \quad \text{if } H \leq L_z, \\
        \pdx{}\left[K_s L_z \pdx{H} + \frac{\sqrt{S_x}}{n_s} \left(H-L_z\right)^k\right] + r & \quad \text{if } H > L_z.
    \end{cases}
\end{equation}
Here, $x$ is distance from the river, located at $x = 0$, $t$ is time; $L_z$ is the soil depth; $K_s$ is the saturated hydraulic conductivity; $S_x$ is the elevation gradient; $r$ is the rainfall; $n_s$ is Manning's roughness coefficient; $k=\frac{5}{3}$ is an exponent from Manning's law \citep{manning1891flow}, and $f(x)$ characterises the mean drainable porosity, which depends on the $x$-dependent initial saturation of the soil. 
As demonstrated in our last paper, the above 1D model gives predictions that are consistent with a full 2D hillslope model under certain conditions where the flow predominantly flows in the $x$ direction towards the river.

\begin{figure}
    \centering
    \def\svgwidth{0.9\linewidth}
    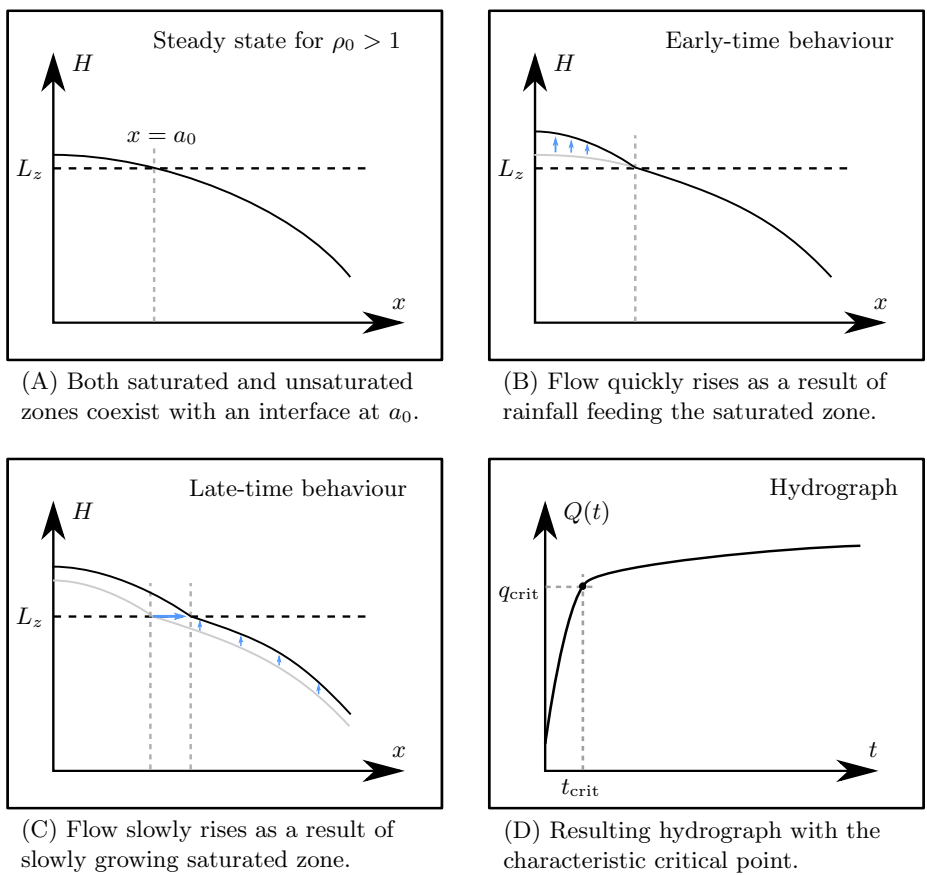
    \caption{Early- and late-time behaviour predicted by the physical model. $H$ is the total height of groundwater and surface flow ($H=L_z$ corresponds to the groundwater reaching the surface). The surface water height was magnified to better illustrate its growth.}
    \label{fig:early_late_time_behaviour}
\end{figure}

Most crucially, numerical and analytical analysis of the above model demonstrates that solutions exhibit distinct early- and late-time behaviours---these are illustrated in \cref{fig:early_late_time_behaviour}. As seen in insets (A) and (B), at early times, the river flow rises rapidly as a result of surface water accumulating over a part of the catchment that is initially fully saturated. However as seen in insets (C) and (D), in the late-time regime, the river flow rises slowly as a result of the rising groundwater, thus leading to the growth of the seepage zone.

The distinction between the two phases explained above is a consequence of the large difference between the short characteristic time of surface flow and the long characteristic time of groundwater flow. Thus, within the short timescale, the surface water profile reaches a quasi-static state, which then continues to adapt to the slowly rising groundwater during the subsequent long timescale.

As shown in \cite{paper3}, it is possible to derive asymptotic approximations for the critical time, $\tsat$, where the quasi-static state of the surface flow is reached; analysis can also yield approximations of the total river inflow value, $\qsat$, at this moment.These key quantities, illustrated in \cref{fig:1D_hillslope}(D), are expressed in terms of the catchment's physical parameters as follows:
\begin{subequations} \label{eq:tsatqsat}
	\begin{equation}
	    \label{eq:tsat_dimensional}
	    \tsat = \frac{L_z}{r}\left[\frac{S_x^{1/2}K_s n}{L_z^{k-1}} \left(\frac{L_x r}{K_s S_x L_z} - \frac{r}{r_0}\right)\right]^{1/k},
	\end{equation}
	\begin{equation}
	    \label{eq:qsat_dimensional}
	    \qsat = \underbrace{K_s S_x L_z}_{\text{groundwater flow}} + \underbrace{r L_x \left(1 - \frac{K_s S_x L_z}{r_0 L_x}\right)}_{\text{surface flow}},
	\end{equation}
\end{subequations}

Therefore, what we are interested in is whether conceptual models can also reproduce these key quantities of saturation time and flow, when modelling the same simplified hillslopes considered above. We expect that in order to accurately reconstruct the benchmark hydrograph, the conceptual models should also be characterised by distinct timescales. It is then possible to ask whether the conceptual model, when studied at the thresholds of the timescales observed, also demonstrate agreement with \eqref{eq:tsat_dimensional} and \eqref{eq:qsat_dimensional}. Do the conceptual models produce analogues of \eqref{eq:tsatqsat} and if so, what are the parametric dependencies (for the conceptual model)? 

\section{A continuous-time Probability Distributed Model (PDM)}
\label{sec:PDM}

\subsection{Probability Distributed Model for the storage capacity}

\noindent Since the properties of real-world catchments are highly heterogeneous, instead of parameterising the entire catchment with a single numerical parameter (\emph{e.g.} a fixed infiltration capacity), many conceptual models represent such properties with a distribution of parameter values. An early example is the Stanford Watershed Model \citep{crawford1966digital}, which used a probability-distributed infiltration capacity (cf. also \citealt{dawdy1978users}, \citealt{moore1981distribution}, and \citealt{dunne1982models}). Introducing this infiltration capacity to the models allows for the generation of surface flow even for small rainfall, making predictions more consistent with the flow data.

Here, we briefly describe a Probability Distributed Model (PDM) for the storage capacity, introduced by \cite{moore1985probability}, which forms the basis for the conceptual model analysed in this paper. We assume that the catchment is made up of many storage elements, each one characterised by a moisture storage capacity or store depth, $c$ (units of [L]). For each such store, $c$ then denotes the total volume of water that can be absorbed by the soil per surface area. In order to model the variation of storage over the catchment area, we posit a cumulative distribution function, $F(\tilde{c})$, which describes the fraction of the basin that has $c \leq \tilde{c}$. Moore suggested that a reflected power distribution function can be used for $F$, and hence,
\begin{equation}
    \label{eq:prob_distribution}
    \text{Prob}(c \leq \tilde{c}) = F(\tilde{c}) = 1-\left(1-\frac{\tilde{c}}{c_{\max}}\right)^b.
\end{equation}
Here, $c_{\max}$ is the maximum store capacity of the basin, and $b > 0$ is a parameter characterising the shape of the distribution.

At each point in time, we let $c^*(t)$ be a measure of the current level of moisture storage in the catchment. \cite{moore1985probability} refers to $c^*$ as the \emph{critical capacity}, \emph{i.e.} the level at which all stores with $c < c^*$ are full and will contribute to the direct runoff.

\begin{figure}
    \centering
    \def\svgwidth{\linewidth}
    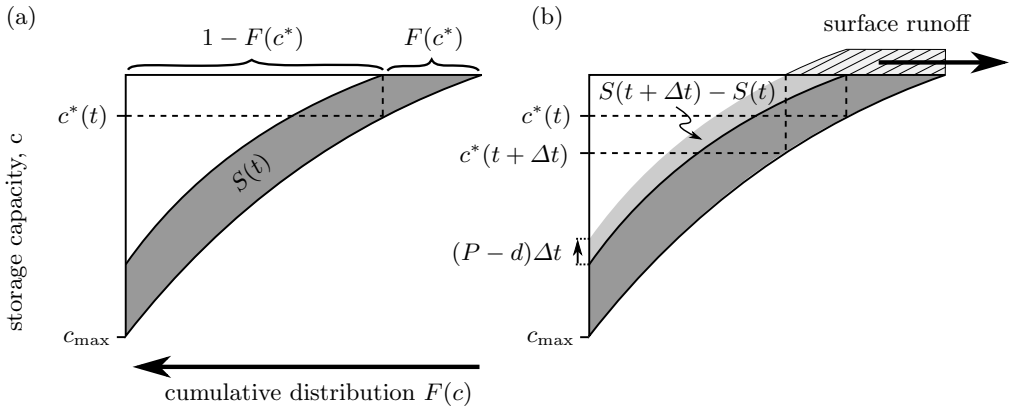
    \caption{Surface runoff production in the PDM model by Moore et al.~\cite{moore1985probability}. (a) Initially, the moisture level is $c^*(t)$, which corresponds to the fraction $F(c^*)$ of the basin being fully saturated. (b) If due to rainfall, the soil moisture level $c^*$ is rising, surface runoff over the saturated part of the basin is generated.}
    \label{fig:PDM_surface_runoff_production}
\end{figure}

The total basin water storage is $S(t)$ (in [L]). It can be derived by considering those storages that are non-full at a given $c$, \emph{i.e.} $1 - F(c)$, and integrating from $c = 0$ to a depth of $c^*$. Thus,
\begin{equation}
    \label{eq:S_from_C}
    S(t)=\int_0^{c^*(t)}\Big(1-F(c)\Big) \, \mathrm{d}c=S_{\max}\left(1-\left(1-\frac{c^*(t)}{c_{\max}}\right)^{b+1}\right),
\end{equation}
where the maximum basin storage is $S_{\max} = \frac{c_{\max}}{b+1}$. There is a geometrical interpretation of \eqref{eq:S_from_C}, as shown in \cref{fig:PDM_surface_runoff_production}(a). The shaded area represents a constant moisture storage level $c^*$, while the white area represents the remaining part of the storage capacity. As highlighted in the figure, the fraction of the basin which is fully saturated (\emph{i.e.} in which $c^*$ is greater than the storage capacity) is represented by $F(c^*)$, and similarly, the unsaturated part is given by $1-F(c^*)$. It can be verified that the shaded area between the two curves is equal to \eqref{eq:S_from_C}, providing a geometrically intuitive interpretation of each store's level rising by $c^*$ from its maximal capacity.

The role of the water storage in the production of fast surface flow in the system is as follows. From time $t$ to time $t + \Delta t$, the water balance requires
\begin{equation} \label{eq:balance}
\underbrace{(P(t) - d(t))\Delta t}_{\text{precipitation} - \text{recharge}}
 = \underbrace{u_f(t)\Delta t}_{\text{surface runoff}} + \underbrace{S(t + \Delta t) - S(t)}_{\text{increase in moisture storage}}.
\end{equation}
Here, $P(t)$ is the precipitation rate, $d(t)$ is groundwater recharge, which refers to the transfer of water to the slow groundwater system, and $u_f(t)$ is surface runoff, \emph{i.e.} the transfer of water to the fast surface water system. All three quantities are expressed in [L/T]. In \cref{fig:PDM_surface_runoff_production}(b), the initial soil moisture storage $S(t)$ is illustrated by the dark gray region. As a result of precipitation, the moisture storage increases as illustrated by the light gray region, and the surface runoff is generated over the saturated area, represented by the hatched region. The key to the probability-distributed storage is that even a small increase in $c^*$ can cause water in some parts of the stores to exceed their capacity and produce runoff.

The above-introduced PDM for storage capacity has been used in many conceptual models, including distributed models such as Grid Model by \cite{bell1998grid} and the Grid-to-Grid model by \cite{bell2007development}. \cite{bell1999elevation} formulated a general form of lumped models based on this PDM, for which a six-parameter version was calibrated to real-world data by \cite{lamb1999calibration}.

\subsection{Six-parameter PDM implementation}

\noindent A notable example of an application of a PDM is the six-parameter model by~\cite{lamb1999calibration}. The discrete formulation of this model from the original paper is presented in \cref{app:discrete_model}. Here, however, we present a more general continuous version obtained by taking $\Delta t \rightarrow 0$. The continuum formulation is a more useful form to highlight differences between the PDM and our physical benchmark model, presented later.

\begin{figure}
    \centering
    \def\svgwidth{\linewidth}
    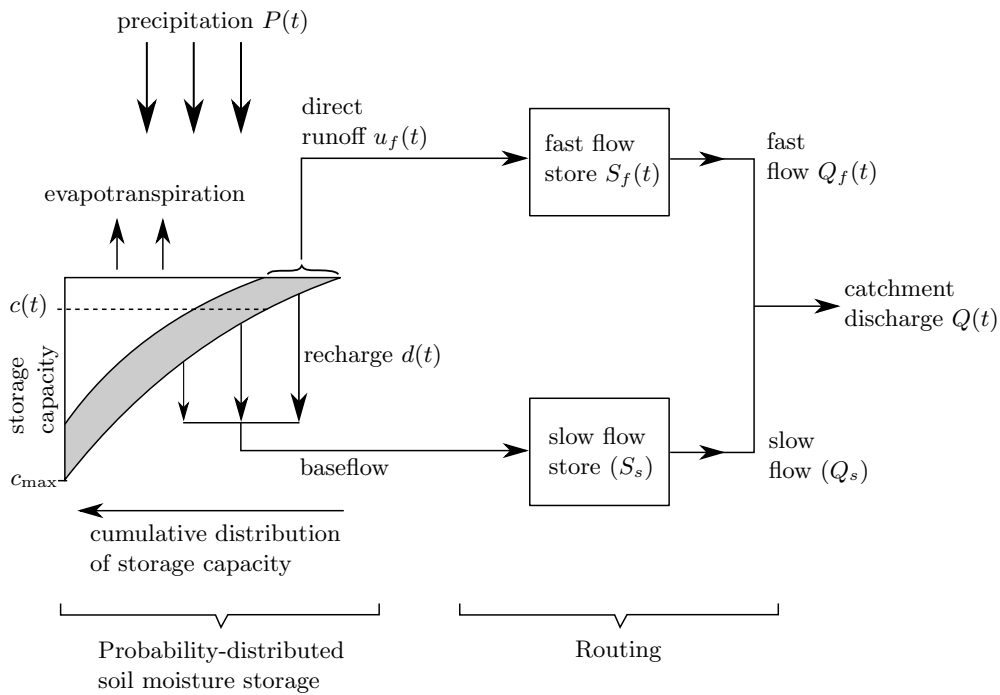
    \vspace*{0.1cm}
    \caption{Structure of the six-parameter Probability-Distributed Model~\citep{lamb1999calibration}.}
    \label{fig:PDM_scheme}
\end{figure}

The lumped model essentially consists of three unknowns: (i) the current critical moisture capacity, $c^*(t)$, which in effect determines the moisture water storage, $S(t)$; (ii) the \textit{slow storage}, $S_s(t)$, representing the slowly moving groundwater; and (iii) the \textit{fast storage}, $S_f(t)$, representing the fast surface flow. The interaction between these models is illustrated in \cref{fig:PDM_scheme} and explained below. 

The three key unknowns are governed by first-order differential equations. First, the critical moisture capacity is
\begin{equation}
  \label{eq:c_ODE}
  \dt{c^*(t)}=\begin{cases}
  0 & \quad\text{for }c^*(t)=c_{\max}\text{ and }P(t)>d(t),\\
  P(t)-d(t) & \quad\text{otherwise}.\\
  \end{cases}
\end{equation}
The top condition imposes the condition that, if $c^*$ reaches the maximum moisture store capacity ($c_{\max}$), but the precipitation still outweighs the drainage, then the store capacity does not rise further. In our simplified scenario, we consider $P$ to be the total precipitation over the hillslope, expressed as $P(t)=r(t)L_x$. Also, as in the case of our physical benchmark, we do not include the effect of evapotranspiration.

The drainage, $d$, is dependent on the total soil moisture, $S(c^*)$, as
\begin{equation}
    \label{eq:drainage}
    d(t)=\max\left(0, \frac{S(c^*(t))-s_t}{k_g}\right),
\end{equation}
where $k_g$ and $s_t$ are constant model's parameters. For $S<s_t$, no more drainage is produced. 

Next, the equations governing the slow and fast storages are:
\begin{subequations}
    \label{eq:sf_ss_ODE}
    \begin{align}
        \label{eq:ss_ODE}
        \dt{S_s(t)} &=d(t)-Q_s(t), \\
        \label{eq:sf_ODE}
        \dt{S_f(t)} &=u_f(t)-Q_f(t).    
    \end{align}
\end{subequations}
Note that the groundwater recharge, $d(t)$, is feeding the slow storage, while surface runoff, $u_f(t)$, is feeding the fast storage. The outflow from both of these storages, $Q_f(t)$ and $Q_s(t)$, contributes to the total catchment discharge $Q(t)$.

The surface runoff $u_f(t)$ is produced by all water that is not absorbed by the soil and is transferred to the groundwater store. From a water balance, we have
\begin{equation} \label{eq:uf}
    u_f(t)= P(t)-d(t) - \dd{S}{t}\biggr\rvert_{c = c^*(t)} = P(t)-d(t)- \left(1-\frac{c^*(t)}{c_\text{max}}\right)^b\dt{c^*(t)}.
\end{equation}
The above formula was explained by \eqref{eq:balance} and uses \eqref{eq:S_from_C}.

Lastly, the surface flow $Q_f(t)$ and groundwater flow $Q_s(t)$ are modelled by linear and exponential stores, \emph{i.e.}
\begin{equation}
  \label{eq:Qf_Qs_Q}
    Q_f(t)=\frac{S_f(t)}{k_f} \quad \text{and} \quad
    Q_s(t)=\exp\left(\frac{S_s(t)}{k_s}\right),
\end{equation}
with the key quantity being the total flow, 
\begin{equation}
    Q(t)=Q_f(t)+Q_s(t).
\end{equation}
Above, $k_f$ and $k_s$ are constant parameters. Note that the linear fast storage has to be non-negative ($S_f\geq 0$), while the exponential groundwater storage can be either positive or negative.

It is convenient to summarise the PDM implementation into a final system. In the situation of constant precipitation, $P(t) = P$, we have
\begin{subequations}
\label{eq:PDM_summary}
\begin{align}
    \label{eq:PDM_summary_a}
   \dd{c^*}{t} &= \begin{cases}
       0, & \text{if $c^* \geq c^*_{\max}$}, \\
       P - d(t), & \text{if $c^* < c^*_{\max}$}, 
   \end{cases} \\
    \label{eq:PDM_summary_b}
   \dd{S_s}{t} &= d(t) - \exp\left(\frac{S_s}{k_s}\right), \\
    \label{eq:PDM_summary_c}
   \dd{S_f}{t} &= \left[P - d(t) - \left(1-\frac{c^*(t)}{c_\text{max}}\right)^b\dt{c^*(t)}\right] - \frac{S_f}{k_f},
\end{align}
\end{subequations}
with $d$ given by \eqref{eq:drainage}. Initial conditions are required at $t = 0$, and these will be specified in \cref{sec:steady}. Notice that our continuum formulation of the PDM implementation by~\cite{lamb1999calibration} involves six parameters: $c_{\max}$, $b$, $k_g$, $k_s$, $k_f$ and $s_t$. These parameters will be fitted to the available training data. The choice of precipitation, $P$, will characterise each particular benchmark scenario, as will the chosen initial condition.

\section{Numerical results and identification of three regimes} \label{sec:regimes}

\subsection{Steady-state configurations} \label{sec:steady}

\noindent As in the case of the physical benchmark described in \cref{sec:physical_benchmark}, we assume that initially the system is in equilibrium for some mean rainfall $P_0$, after which at $t=0$ rainfall $P(t)=P$ will start to fall. Steady-state requires
\begin{equation}
    \label{eq:ic}
    \dt{c^*(t)}=\dt{S_s(t)}=\dt{S_f(t)}=0\,\bigg|_{t=0}\quad\text{for }P(t)=P_0.
\end{equation}
Setting time derivatives in \eqref{eq:PDM_summary_a}--\eqref{eq:PDM_summary_c} to zero with $P(t)=P_0$ gives us three equations for the initial values $c^*(0)$, $S_s(0)$, and $S_f(0)$, namely:
\begin{subequations}
\label{eq:steady_state}
\begin{align}
    \label{eq:steady_state_a}
    0 &= \begin{cases}
       0, & \text{if $c^* \geq c^*_{\max}$}, \\
       P_0 - d, & \text{if $c^* < c^*_{\max}$}, 
    \end{cases} \\
    \label{eq:steady_state_b}
    0 &= d - \exp\left(\frac{S_s}{k_s}\right), \\
    \label{eq:steady_state_c}
    0 &= \left[P_0 - d\right] - \frac{S_f}{k_f}.
\end{align}
\end{subequations}

Since there is an upper limit on the capacity of the soil moisture storage ($c_{\max}$), then the groundwater recharge $d$ given by~\eqref{eq:drainage} with \eqref{eq:S_from_C} cannot exceed a certain threshold value:
\begin{equation}
    d_{\max}=\frac{1}{k_g}\Big(S\left(c_{\max}\right) - s_t\Big)=\frac{1}{k_g}\left(\frac{c_{\max}}{b + 1} - s_t\right).
\end{equation}
Therefore, \eqref{eq:steady_state_a} has two possible solutions depending on whether $P_0$ is lower or greater than $d_{\max}$. If $P_0\geq d_{\max}$, then the soil has to be fully saturated $c=c_{\max}$, and the corresponding drainage is $d=d_{\max}$. If $P_0<d_{\max}$, then the second condition has to be met, $d=P_0$, and the value of $c^*$ can be found by solving \eqref{eq:drainage}. To summarise:
\begin{equation}
    d = \begin{cases}
       d_{\max}, & \text{if $P_0 \geq d_{\max}$}, \\
       P_0, & \text{if $P_0 < d_{\max}$}, 
    \end{cases} \\
\end{equation}
Now, we can use equations \eqref{eq:steady_state_b}--\eqref{eq:steady_state_c} to determine the steady-state value of $S_s$ and $S_f$:
\begin{subequations}
\begin{align}
    S_s &= k_s\ln{d}, \\
    S_f &= k_f\left(P_0 - d\right).
\end{align}
\end{subequations}
Note that if $P_0<d_{\max}$, then $S_f=0$, and so no overland flow initially exists. Note that two types of steady states are possible; with the time-dependent simulations initiated from this state, we will observe multiple possible time-dependent behaviours.

\subsection{Numerical time-dependent solution}

\noindent Time-dependent numerical solutions of the PDM are computed using the finite-difference scheme described in \cref{app:discrete_model}.

\begin{figure}
    \centering
    \includegraphics{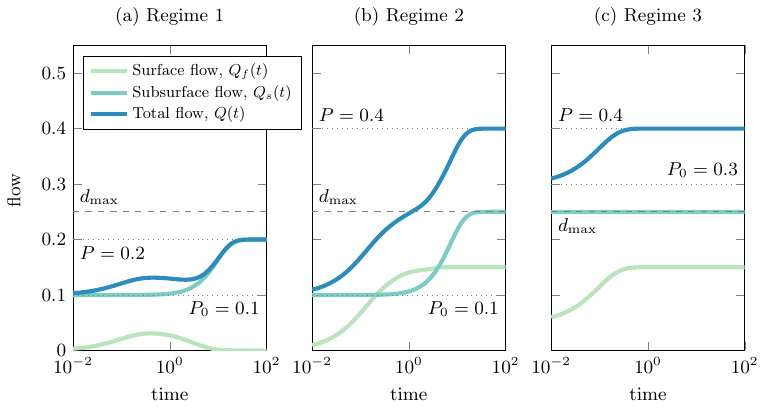}
    \caption{Example hydrographs obtained using PDM for three cases: (a) $P_0<P<d_{\max}$, (b) $P_0<d_{\max}<P$, and (c) $d_{\max}<P_0<P$. The following parameter values were used: $c_{\max} = 1$, $b = 3$, $k_g = 1$, $s_t = 0$, $k_f = 0.1$, $k_s = 1$.}
    \label{fig:PDM_solution_space}
\end{figure}

We can distinguish three qualitatively different regimes in our benchmark scenario. In \emph{Regime 1} ($P_0<P<d_{\max}$), the precipitation never exceeds the groundwater infiltration capacity, and so the soil moisture storage never becomes fully saturated. In \emph{Regime 2} ($P_0<d_{\max}<P$), the soil moisture storage eventually becomes fully saturated. In \emph{Regime 3} ($d_{\max}<P_0<P$), the soil moisture storage is initially fully saturated. Example hydrographs corresponding to each regime are presented in \cref{fig:PDM_solution_space}. We can observe that some of the observed regimes are inconsistent with the behaviour of our physical benchmark, discussed in \cref{sec:physical_benchmark} (and indeed, with physical expectation).

In Regime 1, observe that the initial surface flow rises as a result of increased precipitation accumulating over the saturated part of the soil. However, after soil moisture approaches steady-state, the surface flow decreases asymptotically to zero. In our benchmark model, the surface flow cannot decrease during a constant precipitation $P>P_0$ (by virtue of the geometry of the hillslope). In the PDM, we can experience a local maximum of catchment discharge, which is unrealistic from the point-of-view of the physical model.

Another difference is observed in Regime 2 (\cref{fig:PDM_solution_space}b) where we see the total flow rises twice: the first rise is caused by the rise of the surface flow, and the second rise is caused by the rise of the groundwater flow. In the physical benchmark, we observe only a single period of rapid rise of total flow followed by a late-time behaviour in which the rate of change of total flow decays in time. Also, in the physical benchmark, the groundwater flow stays constant in time.

Turning now to Regime 3 (\cref{fig:PDM_solution_space}c). Here, groundwater flow stays constant, but we do not observe a distinct early- and late-time behaviour---the surface flow just quickly rises to reach a new steady-state value (observed in the physical benchmarks). 

Our goal now is to understand the reason for these qualitative differences, and to further understand how PDM can be used to approximate solutions from a more fundamental physics-based model. 

\section{Asymptotic analysis of the PDM} 

\noindent Our analysis in this section will demonstrate that the PDM exhibits three distinct timescales that can be observed in the simulations. These timescales can then be used as a method for calibrating the control parameters.

\subsection{The three timescales}
\label{sec:timescales}

\noindent Let us consider Regime 3 ($P_0>d_{\max}$) first. From the initial condition imposed in~\eqref{eq:ic}, we have:
\begin{equation}
c^*(0)=c_{\max},\quad S_f(0)=k_f\left(P_0-d_{\max}\right),\quad S_s(0)=k_s\ln\left(d_{\max}\right).
\end{equation}
Thus, the soil moisture storage is fully saturated, while $S_s$ and $S_f$ have the initial value corresponding to subsurface flow, $Q_s=d_{\max}$, and surface flow, $Q_f=P_0-d_{\max}$, respectively.

Now, if at $t=0$ the precipitation rate rises to $P$, the discharge, $d$, does not change since it has already reached its maximal value. As a consequence, the groundwater flow does not change either, staying equal to $Q_s=d_{\max}$. The recharge, however, reaches a new constant value equal to $r=P-d_{\max}$. Consequently, following equations~\eqref{eq:sf_ODE} and \eqref{eq:Qf_Qs_Q}, we have 
\begin{equation}
    \label{eq:saturated_case}
    \dt{S_f(t)}=P-d_{\max}-\frac{S_f(t)}{k_f},
\end{equation}
and thus
\begin{equation}
\label{eq:analytic_solution}
S_f(t)=S_f(0)\e^{-\frac{t}{k_f}}+k_f\left(P-d_{\max}\right)\left(1-\e^{-\frac{t}{k_f}}\right).
\end{equation}
Therefore, in this regime, the following hydrograph results:
\begin{equation}
Q(t)=Q_s(t)+Q_f(t)=d_{\max}+\frac{S_f(t)}{k_f}=P_0 \e^{-\frac{t}{k_f}}+P\left(1-\e^{-\frac{t}{k_f}}\right).
\end{equation}
As expected, the flow rises from an initial value of $Q(0)=P_0$ to a new equilibrium $Q \to P$ as $t\rightarrow\infty$. The dynamics of this process are governed only by a single timescale, $t_f=k_f$, representing the characteristic time of fast storage discharge.

As an alternative to solving the ODEs, the characteristic timescale can also be derived via nondimensionalisation. Re-scaling $t'=k_f t$ and $S_f'(t)=\left(P-d_{\max}\right)S_f(t)$ in the original equation~\eqref{eq:saturated_case} yields $\mathrm{d}{S_f'}/\mathrm{d}t'=1-S_f'$. Since this form does not include any parameters, we see that indeed the solution is characterised by only a single timescale $t_f=k_f$. Thus, the PDM in this case does not allow a reconstruction of the characteristic behaviour of our physical benchmarks, which contain distinct early- and late-time behaviours (see \cref{sec:physical_benchmark}).

However, we can observe additional timescales in the partially unsaturated case, \emph{i.e.} Regimes 1 and 2 ($P_0<d_{\max}$). Here the initial condition~\eqref{eq:ic} gives:
\begin{equation}
    \label{eq:partially_saturated_ic}
    S(0)=s_t+k_g P_0,\qquad S_f(0)=0,\qquad S_s(0)=k_s\ln\left(P_0\right),
\end{equation}
that is, we have no initial surface flow, while both $S$ (or $c^*$) and $S_f$ correspond to the groundwater recharge $d$ and groundwater flow $Q_f$ equal to the initial precipitation $P_0$, respectively. The further flow is governed by ODEs \eqref{eq:c_ODE} and~\eqref{eq:sf_ss_ODE}, and cannot be solved analytically, however, we still can identify the timescales by finding a dimensionless version of these equations. This is demonstrated in detail in \cref{app:three_timescales} following the methodology presented in the fully saturated case.

Three timescales are identified, corresponding to the dynamics of $c^*(t)$, $S_f(t)$, and $S_s(t)$. Respectively, these are
\begin{equation}
\begin{aligned}
t_c = \frac{c_{\max}}{P} &= \frac{\text{soil saturation}}{\text{precipitation}}, \\
t_f =k_f &= \text{fast storage dynamics}, \\
t_s = \frac{k_s}{d_{\max}} &= \frac{\text{groundwater discharge}}{\text{groundwater recharge}} = \text{slow storage dynamics}.
\end{aligned}
\end{equation}
The first timescale is the characteristic time over which the soil becomes fully saturated. The second timescale is the same timescale characterising fast storage dynamics as derived in the fully saturated case (Regime 3). The last timescale characterises the slow storage dynamics and is dependent on both groundwater recharge (parametrised by $d_{\max}$) and the rate of its discharge (parametrised by $k_s$).

\begin{figure}
    \centering
    \includegraphics{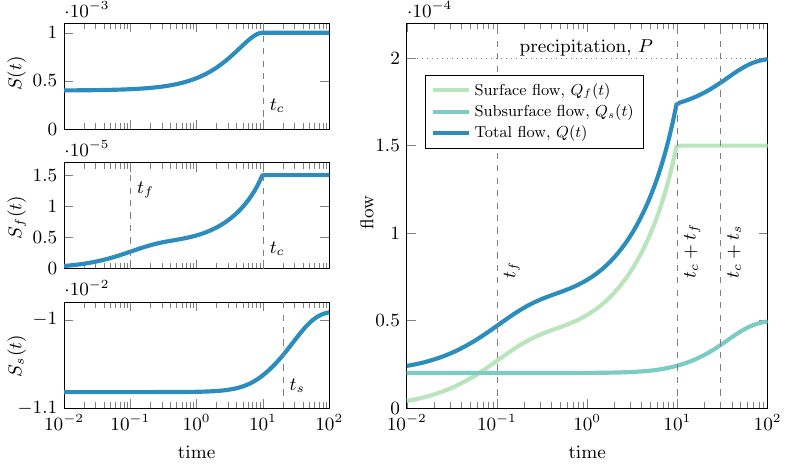}
    \caption{Example PDM simulation with three distinct timescales $t_c<t_f<t_s$. The evolution of each storage with corresponding timescale is presented on the left. The resulting hydrograph is presented on the right.}
    \label{fig:PDM_timescales}
\end{figure}

Depending on the relative size of those terms, we may experience different short-, mid-, and long-time behaviour. Consider, for example, $c_{\max}=0.002$, $b=1$, $k_g=20$, $k_f=0.1$, $k_s=0.001$, $s_t=0$, $P_0=2\cdot 10^{-5}$, and $P=2\cdot 10^{-4}$. The characteristic timescales are then $t_f=0.1$, $t_c=10$, and $t_s=20$, so there is significant separation between them. The numerical solution for this model is presented in \cref{fig:PDM_timescales}.

As expected, the variation of moisture storage $S(t)$ (or $c^*(t)$) takes place around $t=\Oh(t_c)$ (\cref{fig:PDM_timescales}a). Consider firstly the fast storage: we observe a quick rise near $t=\Oh(t_f)$, followed by a slower rise in timescale $t=\Oh(t_c)$ (\cref{fig:PDM_timescales}b). The quick rise is due to the fast store adapting to the increased surface runoff from the initial seepage zone. The longer timescale represents the slow increase of surface runoff caused by the increasingly-full basin, which becomes fully saturated during the moisture-store saturation process. All three timescales can be observed in the hydrograph presented in \cref{fig:PDM_timescales}d. Finally, note that the slow storage evolves at the slowest timescale, $t_s$ (\cref{fig:PDM_timescales}c).

As introduced in \cref{sec:physical_benchmark}, we note that the physical model discussed there includes two timescales, rather than three like in the case of PDM. In essence, this is because in our physical benchmark model, the subsurface flow stays constant; its main impact on the total flow is via its coupling with the surface flow. As we show further in \cref{sec:fitting_and_validation}, PDM can still be fitted to these catchments by setting a low $k_s$ value giving a short $t_s$ timescale, so that no additional long-time behaviour is observed.

\begin{figure}
    \centering
    \def\svgwidth{\linewidth}
    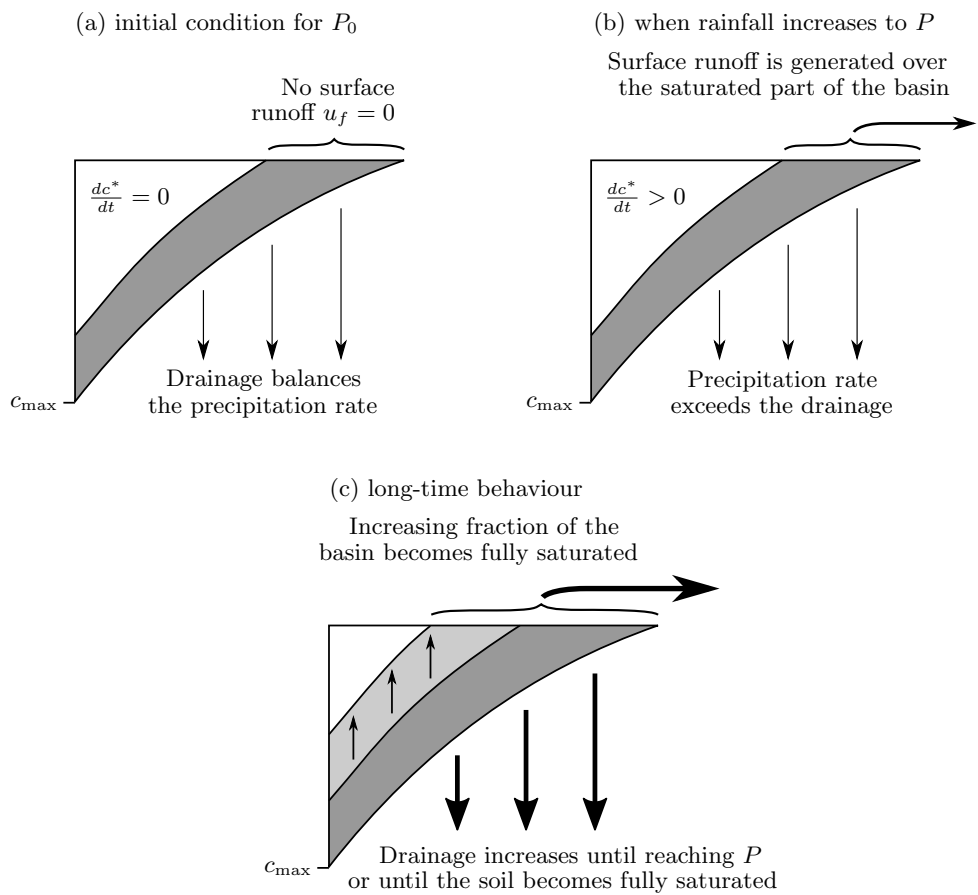
    \caption{The schematic diagram of the response of the PDM soil moisture storage to an intensive rainfall in $P_0<d_{\max}$ scenario.}
    \label{fig:short_long_time_mechanism}
\end{figure}

Nevertheless, the remaining two timescales have the same physical interpretation in both models. Similarly to the benchmark model presented in \cref{fig:early_late_time_behaviour}, in the case of the PDM, part of the catchment is initially fully saturated (\cref{fig:short_long_time_mechanism}a), however, with no surface runoff being generated over the seepage zone. The short timescale in both models corresponds to the precipitation accumulating over this initial seepage zone (\cref{fig:short_long_time_mechanism}b). Similarly, the long timescale in both models is related to the slow rise of soil water content, causing the size of the seepage zone to gradually grow (\cref{fig:short_long_time_mechanism}c).

\subsection{Early-time quantitative  comparison}
\label{sec:early_time_comparison}

\noindent We have already shown that the results of the PDM and the physical model are largely consistent; indeed, both models possess distinct early- and late-time behaviours with similar physical interpretations. In this section, we explore the quantitative properties of the early-time behaviour, as argued in \cref{sec:physical_benchmark}, the early-time dynamics are responsible for the dominant peak flow generated during rainfalls in our benchmark.

In \cref{app:three_timescales}, we have already found that the short-time behaviour is given by a single timescale $t_f = k_f$, where $k_f$ is a fitted catchment-specific parameter. However, according to the physical benchmark model, the corresponding saturation time $\tsat$ is given by~\eqref{eq:tsat_dimensional}. Since $\tsat \propto r^{1/k-1} = r^{-2/5}$, it is not only a function of catchment/hillslope properties, but also decreases with the precipitation rate $P$ (proportional to $r$). As a consequence, if we fit the value of $k_f$ to standard rainfalls, we will overestimate the time during which the river flow rises in the case of very high rainfalls, or underestimate it in the case of small rainfalls.

Similarly, according to~\eqref{eq:tsat_dimensional}, $\tsat$ increases with the mean precipitation $r_0$. This is due to the fact that increasing mean precipitation increases the size of the seepage zone, which the surface flow has to cross in order to reach the channel. As we stated in \cref{sec:timescales}, in the PDM, the corresponding timescale $t_f = k_f$ is constant, \emph{i.e.} independent of $P_0$ (proportional to $r_0$) as well. Therefore, we should expect that the PDM either overestimates the timescale of the early-time rapid increase of river inflow for seasons with low $r_0$ (or $P_0$) value or underestimates this time for seasons with high $r_0$ value.

Now, let us compare the value of the flow, which is reached during this early-time period. This flow corresponds to the quasi-static state of the fast flow store, in which it is already adapted to the surface runoff produced over the seepage zone (within timescale $t_s$), but not impacted by the much slower increase of the saturation zone, propagating with a timescale $t_c \gg t_s$. In \cref{app:quasistatic_flow}, we show that the flow increase $\Delta Q$ for the PDM is given by
\begin{equation}
    \label{eq:dq_pdm}
    \Delta Q_\text{PDM} = (P-P_0) \underbrace{\left[1-\left(1-\frac{s_t+k_gP_0}{s_\text{max}}\right)^\frac{b}{b+1}\right]}_{a_\text{PDM}(P_0)}.
\end{equation}

The flow reached according to the physical benchmark model is given by~\eqref{eq:qsat_dimensional}, which corresponds to the rise of flow when the precipitation rate changes from $P_0=r_0L_x$ to $P=rL_x$ equal to:
\begin{equation}
    \label{eq:dq_physical}
    \Delta Q_\text{benchmark} = (P - P_0) \underbrace{\left(1 - \frac{K_s S_x L_z}{P_0}\right)}_{a_0(P_0)}.
\end{equation}
Note that~\eqref{eq:dq_pdm} and~\eqref{eq:dq_physical} have a similar form. Both functions are expressed as ${\Delta Q=(P - P_0)a(P_0)}$, where $a(P_0)\in[0,1]$ in both cases is a function of catchment parameters and $P_0$. As we explained in \cref{sec:physical_benchmark} following our earlier work, this parameter has a simple interpretation---it is the size of the initial seepage zone.

\begin{figure}
    \centering
    \includegraphics{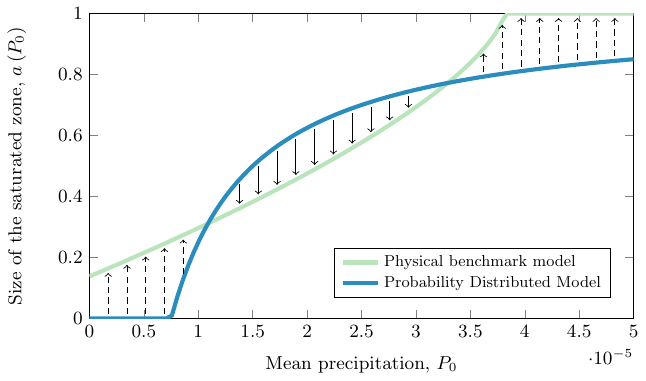}
    \caption{The size of the seepage zone according to the PDM~\eqref{eq:dq_pdm} and physical benchmark model~\eqref{eq:dq_physical}. Due to the opposite curvature of each function, the PDM underestimates $a(t)$ for intermediate values of $P_0$ (solid arrows), and overestimates it for low and high values of $P_0$ (dashed arrows).}
    \label{fig:a_shape}
\end{figure}

In both cases, $a(P_0)$ increases with a mean precipitation $P_0$. However, the dependence on $P_0$ in both cases has a different functional form. The key difference is that $a_0(P_0)$ is a concave function of $P_0$, while $a_\text{PDM}(P_0)$ is convex. Consequently, there can be at most two $P_0$ values, in which $\Delta Q_\text{benchmark}(P_0)=a_\text{PDM}(P_0)$. Therefore, we deduce that the PDM model can predict the flow accurately only around two $P_0$ values, with peak flow being underestimated between them, and overestimated otherwise (see \cref{fig:a_shape}).

Since $P_0$ (and resulting soil moisture) depends on the season, the PDM cannot be fully consistent with the physical benchmark over any conditions, ranging from low $P_0$ values in dry months to high $P$ in wet months, unless we fit PDM parameters for each season separately. We further explore this dependence in \cref{sec:P0_dependence}.

Apart from the seasonal variation of mean precipitation/evaporation rates, this effect can possibly be observed for the catchments for which these properties change systematically, \emph{e.g.} due to climate change. Increasing water circulation can increase $P_0$, causing models fitted to the historical data to be no longer accurate.

Comparison of the early-time behaviour predicted by these two models is useful for two primary reasons. First, the comparison has allowed us to identify the above sources of inaccuracy. Second, it allows us to find how the PDM-fitted parameters are related to physical catchment properties. Here, the transition from short- to long-time behaviour on the hydrographs is matched when
\begin{equation}
    k_f\propto \tsat = \frac{L_z}{r}\left[\frac{S_x^{1/2}K_s n}{L_z^{k-1}} \left(\frac{L_x r}{K_s S_x L_z} - \frac{r}{r_0}\right)\right]^{1/k},
\end{equation}
and
\begin{equation}
    a = 1-\left(1-\frac{s_t+k_gP_0}{s_\text{max}}\right)^\frac{b}{b+1} = 1-\frac{K_s S_x L_z}{P_0}.
\end{equation}
Potentially, additional relations similar to the above can be developed by investigating other benchmark scenarios; for example, one can examine the drying process after an intensive rainfall, as used by~\cite{lamb1999calibration} to calibrate $k_f$ and $k_s$.

\section{PDM fitting and validation}
\label{sec:fitting_and_validation}

\noindent So far, we have identified key qualitative differences between the PDM and our physical benchmark model; this has raised two important questions: (i) can PDM reproduce a hydrograph given by the physical benchmark for a given set of parameters? And (ii) can this fitted model be reliable when applied to different conditions? These conditions may include different values of mean precipitation $P_0$, which affects the initial soil saturation, and different values of peak precipitation $P$ during a rainfall.

To address these questions, we perform three numerical experiments, summarised in this section. The implementation of both models and the numerical experiments is available in a software repository \citep{github_rr}.

\subsection{Comparison of flow components}

\noindent To answer the first question posed above, we construct a hydrograph using the numerical implementation of our physical benchmark model. We used the default parameters presented in Table 1 of \citep{paper3}. For such parameters, the simulation time was extended to seven days so as to allow the hydrograph unto reach (sufficiently near to) the new steady state. Then, we fit a PDM model by finding parameters for which the mean square error of the total flow is the smallest. For minimisation, we used the derivative-free Nelder-Mead method implemented in the \texttt{fminsearch} function in Matlab.

\begin{figure}
    \centering
    \includegraphics{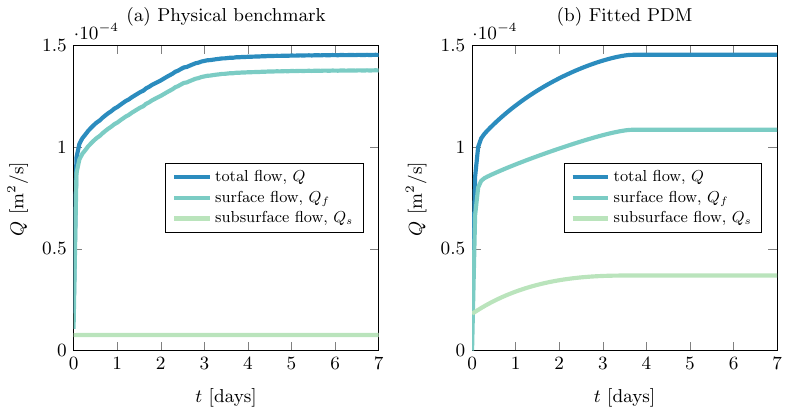}
    \caption{Comparison between the flow components predicted by (a) the physical benchmark and (b) the PDM fitted to the physical hydrograph.}
    \label{fig:Validation_components}
\end{figure}

The result of the fitting is presented in \cref{fig:Validation_components}. Interestingly, despite qualitative differences between the models, the PDM model accurately reproduces the total flow, $Q$, generated by the benchmark model over the entire simulated time period. We can observe the characteristic early-time rapid increase in flow (within approximately 1 hour) caused by the rainfall accumulated over the initial seepage zone. It is followed by a much slower late-time flow rise, eventually reaching a new steady state after a few days.

However, a significant difference can be noticed when decomposing the total flow into its slow and fast components ($Q_s$ and $Q_f$). As discussed in the analysis of Regime 2 in \cref{sec:regimes}, the groundwater flow is increasing in time in the case of PDM, even though it should be constant according to the physical benchmark. In general, as discussed in \cref{sec:timescales}, the groundwater flow introduces an additional (third) timescale. However, in the case of the fitted PDM model, this timescale overlaps with the timescale of the moisture storage. Therefore, the total flow remains consistent for both compared models.

\subsection{PDM accuracy versus mean rainfall \texorpdfstring{$P_0$}{}}
\label{sec:P0_dependence}

\noindent In \cref{sec:early_time_comparison}, we observed that the critical flow reached in early-time increases with the mean rainfall, $P_0$. However, the critical time is predicted by different functional equations for the PDM \eqref{eq:dq_pdm}, as compared to the physical model \eqref{eq:dq_physical}. We concluded that PDM overestimates the flow for high and low $P_0$ values, while the model underestimates the flow for intermediate values.

We wish to demonstrate the above fact more carefully via a numerical experiment. First, we construct five physical hydrographs for different $P_0$ values, corresponding to $a_0=0.20$, $0.35$, $0.5$, $0.65$, and $0.8$, while keeping all other parameters constant. Each hydrograph represents the total river inflow $Q$ per unit river length during a 24-hour intensive rainfall $P=rL_x=1.4538\cdot 10^{-4}~\mathrm{\frac{m^2}{s}}$. Then, we fitted the PDM to match the second and fourth hydrographs. The fitting was obtained by minimising the mean square error of the total flow, averaged over both hydrographs.

\begin{figure}
    \centering
    \includegraphics{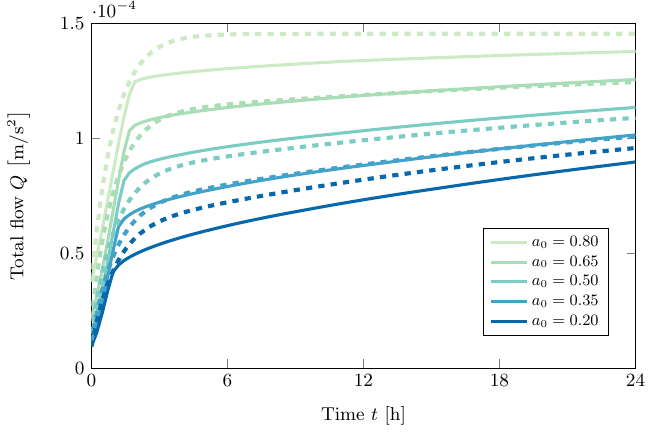}
    \caption{The comparison of the physical benchmark hydrographs (solid lines) and the hydrographs predicted by the PDM (dashed lines) for different mean precipitation $P_0$ values is shown. The PDM was fitted to the $a_0=0.35$ and $a_0=0.65$ hydrographs. The fitted parameters were as follows: $c_{\max}=0.0292$, $b=5.9983$, $k_g=117.1743$, $s_t=7.2155\cdot 10^{-4}$, $k_f=1.0321$, and $k_s=0$.}
    \label{fig:r0_dependence}
\end{figure}

The fitted PDM was then used to reconstruct the shape of all five hydrographs. The results are presented in \cref{fig:r0_dependence}. In the case of the training hydrographs ($a_0=0.35$ and $0.65$), except for a different behaviour around the saturation point, the PDM predicts the flow values properly throughout most of the simulation. However, as we anticipated, the PDM underestimates the flow for the intermediate scenario ($a_0=0.5$) by approximately $4-6\%$, and overestimates it for both high and low $P_0$ values by approximately $8-10\%$.

The difference in the behaviour around the saturation point is mostly due to representing the surface with a linear store, which smoothens the hydrograph compared to a kinematic approximation of Saint Venant equations used in the physical benchmark model, resulting in the characteristic single breaking point at $\tsat$.

Finally, in \cref{fig:r0_dependence}, we can see that the critical time $\tsat$ decreases with $r_0$ (and $a_0$) according to the physical model, as in \eqref{eq:tsat_dimensional}. At the same time, the characteristic time of the early-time growth in the case of the PDM stays approximately the same, as we also postulated in \cref{sec:early_time_comparison}.

\subsection{PDM accuracy vs peak rainfall \texorpdfstring{$P$}{}}
\label{sec:P_dependence}

\noindent Apart from the critical flow, we also observed a difference in behaviour in the predicted critical flow, \emph{i.e.} the time of the early fast increase in the surface flow speed. The characteristic timescale of this process in the PDM model ($t_f$) is only dependent on the constant $k_f$ parameter. On the contrary, the critical flow in the benchmark model depends on multiple physical parameters, including the mean and current precipitation rate, as given by \eqref{eq:tsat_dimensional}.

\begin{figure}
    \centering
    \includegraphics{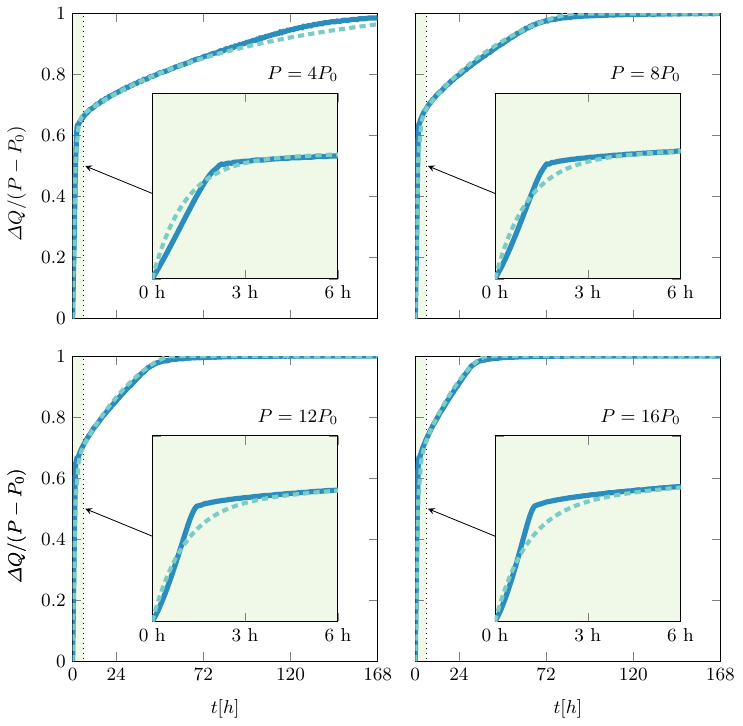}
    \caption{Comparison between PDM predictions (dashed green line) and physical benchmark (solid blue line) for four different precipitation rates. The fitted parameters are $c_{\max}=0.0199$, $b=1.4793$, $k_g=75.3145$, $s_t=0.0053$, $k_f=1.0422$ and $k_s=5.1087\cdot 10^{-6}$. Small lime graphs are a zoomed version of the main graph for the first 6 hours.}
    \label{fig:Validation_set}
\end{figure}

To demonstrate this, we fitted all six parameters of the PDM to match the four physical hydrographs obtained for $P=2P_0$, $P=4P_0$, $P=6P_0$, and $P=8P_0$ (with constant $P_0$), while keeping all other parameters at their default values. The fitting was obtained by minimising the mean square error of the total flow, averaged over all four hydrographs. A sufficiently long (week-long) training hydrograph was used to match the characteristic length of all timescales across both models. Then, the quality of fit was assessed by comparing the obtained hydrographs for scenarios within the training set ($P=4P_0$ and $P=8P_0$) and for more extreme rainfalls ($P=12P_0$ and $P=16P_0$). The results are presented in \cref{fig:Validation_set}.

As predicted in \cref{sec:early_time_comparison}, the duration of the early-time rapid rise of river flow differs between the two compared models. It decreases with the precipitation rate in the physical model but stays constant in the PDM. This causes the PDM to underestimate the rate at which the river flow is growing during short intensive rainfall. However, this inconsistency in the early-time behaviour is slowly eliminated in the late-time behaviour.

\section{Conclusions}

\label{sec:conclusions}

\noindent In this study, we provided a detailed qualitative and quantitative comparison of the Probability-Distributed Model (PDM) and a physical benchmark model corresponding to a simplified hillslope. There are some similarities in the construction of both models, such as separate surface and subsurface flow, each of which is characterised by a different timescale. The introduction of the probability-distributed soil moisture storage allows us to take into account that different parts of the catchment can absorb different amounts of water. The same situation can be observed in the physical benchmark model, in which the areas located further from the river have a lower level of groundwater, so they can absorb more rainfall before becoming saturated. Due to these similarities, the PDM can successfully predict the hydrograph generated with a physical model.

We also observed some fundamental differences between these two models. Firstly, the conceptual PDM model, in its general form, includes some regimes that are unrealistic from a physical point of view. Only when certain conditions are met ($P_0<d_{\max}<P$), it allows us to reconstruct physical hydrographs. Even in this particular regime, the subsurface flow component introduces an additional timescale $t_s$, which does not appear in the physical benchmark solution.

Secondly, we observed that the dependence of the peak flow on the mean rainfall $P_0$ is different in the PDM and the physical benchmark model. As a consequence, it is not possible to match PDM parameters to accurately reconstruct peak flow for all seasons if they are characterised by different groundwater levels. Finally, the PDM underestimates the rate at which the catchment discharge initially rises in the case of short, intensive rainfalls. All identified similarities and differences between these two models are summarized in \cref{tab:summary}.

\begin{table}
    \centering
    \renewcommand{\arraystretch}{1.5}
    \begin{tabular}{{p{2.8cm}p{5cm}p{5cm}}}
        \textsc{Model's component} & \textsc{Physical model} & \textsc{PDM benchmark} \\
        \hline
        
        Soil capacity & It increases with the distance from the river since it is dependent on the initial depth of the groundwater below the surface and mean porosity. & It is given by the reflected power probability distribution $F(c)$. \\
        
        Surface runoff & It is generated by the rainfall accumulating over the seepage zone where $H=L_z$ (saturation-excess flow). & It is generated by rainfall accumulating over the seepage zone where $c>c^*$. \\
        
        Seepage zone size & It increases with the mean precipitation, as given by the concave function \eqref{eq:dq_physical}. & It increases with the mean precipitation, as given by the convex function \eqref{eq:dq_pdm}. \\
        
        \makecell[l]{Seepage zone \\ growth} & It grows as a result of water being accumulated in the soil with a characteristic timescale $T_0=\frac{L_x}{K_sS_x}$. & It grows as a result of water being accumulated in the soil with a characteristic timescale $t_c=\frac{c_{\max}}{P}$. \\
        
        Surface flow & It is given by the Saint-Venant equation with a characteristic timescale $\mu^{1/k}T_0\ll T_0$. & It is given by a linear store model with a characteristic timescale ${t_f=k_f\ll\frac{c_{\max}}{P}}$. \\
        
        Groundwater flow & It stays constant if a seepage zone exists around the channel. & It is given by the exponential store model with an additional characteristic timescale $t_s=\frac{k_s}{d_{\max}}$.\\
        
        Critical point & Critical flow is proportional to the rainfall over the seepage zone, and critical time decreases with the precipitation rate. & Critical flow is proportional to the rainfall over the seepage zone, and critical time depends only on the fitted parameters.\\
    \end{tabular}
    \caption{Summary of similarities and differences between our physical benchmark and PDM.}
    \label{tab:summary}
\end{table}

The approach presented here does not seem to be common in hydrology yet, we think it can help to improve existing methodologies. Rainfall-runoff models are sometimes criticised for not being able to pass through a rigorous cross-validation~\citep{beven2018hypothesis,beven2019make}, \emph{i.e.} there is no guarantee that they will perform well in situations not represented in the training data. \cite{kirchner2006getting} argued that advancing the science of hydrology requires developing not only models which match with the available data, but models that are theoretically justified.

We argue that fundamental studies, such as this one, can contribute to the evaluation of rainfall-runoff models beyond what can be learned from numerical testing alone. Not only do they enable the assessment of model accuracy based on predicted hydrographs, but they also allow us to compare the fundamental assumptions and structures of these models. Additionally, they allow us to predict how the models would perform in different limiting scenarios, such as extreme rainfall events, which are often underrepresented in the available data.

This new perspective can be used to develop or modify conceptual models in order to better align with the possible behaviour of the physical system over the entire range of the model's parameters for a given benchmark scenario. The next essential question is whether these modified models would further increase their accuracy and reliability when applied to real-world data as well. We leave this as an open problem for future studies.

\mbox{}\par
\noindent \textbf{Acknowledgements.} We thank Sean Longfield (Environmental Agency) for useful discussions, and for motivating this work via the 7th Integrative Think Tank hosted by the Statistical and Applied Mathematics CDT at Bath (SAMBa). We also thank Thomas Kjeldsen, Tristan Pryer, Keith Beven and Simon Dadson for insightful discussions. Piotr Morawiecki is supported by a scholarship from the EPSRC Centre for Doctoral Training in Statistical Applied Mathematics at Bath (SAMBa), under the project EP/S022945/1.

\bibliographystyle{plainnat}
\bibliography{bibliography}

\appendix


\section{Discrete version of the six-parameter PDM}
\label{app:discrete_model}

\noindent Here, we formulate the discrete version of the six-parameter PDM as described by \cite{lamb1999calibration}, which we used to develop the continuum version presented in equations~\eqref{eq:c_ODE}--\eqref{eq:Qf_Qs_Q}. This formulation allows us to compute the values of $S$ (basic storage), $S_f$ (fast flow storage), and $S_s$ (slow flow storage) at discrete time steps, $t = t^i$, with $i=1,2,\ldots,N$. Although not all the explicit equations presented in this appendix are not directly provided in the original paper by \cite{lamb1999calibration}, we have reconstructed them based on the information presented in the paper, as well as the original formulation by \cite{moore1985probability}.

At each time step, with $\Delta t = t^{i+1} - t^i$, the soil moisture is increased by $P^i\Delta t$ as a result of precipitation and decreased by $d^i\Delta t$ due to groundwater drainage. However, the moisture storage cannot exceed its maximum capacity, $c_{\max}$. Therefore, we have:
\begin{equation}
    \label{eq:discrete_c_update}
    c^{i+1}=\min\left(c_{\max},c^i+\Pi^i\Delta t\right), \qquad \text{where }\Pi^i \equiv P^i-d^i.
\end{equation}
It is assumed that the groundwater recharge, $d^i$, is given by a linear storage model:
\begin{equation}
    d^i=\max\left(0, \frac{S^i-s_t}{k_g}\right),
\end{equation}
which is zero if the current storage does not exceed some given constant threshold, $s_t$. 

The resultant surface runoff, $u_f^i$, is equal to the amount of precipitation minus drainage ($\Pi^i$), that cannot be absorbed into the basin, \emph{i.e.}
\begin{equation}
u_f^i = \Pi^i - \frac{S^{i+1} - S^i}{\Delta t},
\end{equation}
The assumed form of $S^i$, which depends on $c^i$, is given in \eqref{eq:S_from_C} or alternatively the power distribution function given in eqn~(28) of \citet{moore1985probability}. Then 
\begin{equation}
    u_f^i = \Pi^i + \frac{S_{\max}}{\Delta t} \left[\left(1-\frac{c^{i+1}}{c_{\max}}\right)^{b+1}-\left(1-\frac{c^i}{c_{\max}}\right)^{b+1}\right],
\end{equation}
where $S_{\max} = c_{\max}/(b+1)$, and $b$ is an additional fitted parameter. 

It remains to specify the governing equations for the fast store, $S_f$, and slow store, $S_s$:  
\begin{equation}
    S_f^{i+1}=S_f^i+(u_f^i-Q_f^i)\Delta t \qquad \text{and} \qquad
    S_s^{i+1}=S_s^i+(d^i-Q_s^i)\Delta t.
\end{equation}
Above, the discharges, $Q_f$ and $Q_s$, are modelled as linear and exponential stores, respectively. That is, 
\begin{equation}
    Q_f^i=\frac{S_f^i}{k_f} \qquad \text{and} \qquad 
    Q_s^i=\exp\left(\frac{S_s^i}{k_s}\right),
    \end{equation}
where again $k_f$ and $k_s$ are fitted parameters. Hence, the total discharge is $Q^i=Q_f^i+Q_s^i$.

The reader will confirm there are six parameters, $\{ c_{\max}, b,\, s_t, \, k_g, \, k_f, \, k_s\}$. The above discrete form of the model is used to run the numerical simulations of the PDM, presented in this paper. The numerical code implementing these equations is available in our GitHub repository~\citep{github_rr}.

\section{Characteristic timescales in the partially saturated case}
\label{app:three_timescales}

\noindent In this appendix, we demonstrate how nondimensionalisation of the governing equations can be used to identify three characteristic timescales of the PDM, namely:
\begin{equation}
t_c=\frac{c_{\max}}{P},\quad t_f=k_f,\quad\text{and}\quad t_s=\frac{k_s}{d_{\max}}
\end{equation}

Here, we consider the $P_0<d_{\max}$ case (Regimes 1 and 2); the complementary one (Regime 3) is discussed in detail in~\cref{sec:timescales}. We will investigate two phases: the partially saturated phase when $c^*<c_{\max}$ and the fully saturated phase when $c^*=c_{\max}$. Note that the second phase, when $c^*$ reaches its maximum value, only occurs when $P>d_{\max}$ (\emph{i.e.} in Regime 2).

\subsection{Partially saturated phase}

Following the continuous model formulation given in equations~\eqref{eq:c_ODE}-\eqref{eq:Qf_Qs_Q}, as long as $c^*<c_{\max}$ (\emph{i.e.} before the soil moisture store becomes fully saturated), the dynamics of the system is described by the following three ODEs:
\begin{subequations}
    \begin{equation}
        \label{eq:dim_c}
        \dt{c^*(t)}=P-d(t)
    \end{equation}
    \begin{equation}
        \label{eq:dim_sf}
        \dt{S_f(t)}=\Big(P-d(t)\Big)\left[1-\left(1-\frac{c^*(t)}{c_\text{max}}\right)^b\right]-\frac{S_f(t)}{k_f}
    \end{equation}
    \begin{equation}
        \label{eq:dim_ss}
        \dt{S_s(t)}=d(t)-\exp\left(\frac{S_s(t)}{k_s}\right)
    \end{equation}
\end{subequations}

We can identify the timescales characterising individual equations through nondimensionalisation, as it was demonstrated for Regime 3 in \cref{sec:timescales}. Let us consider only situations when $c^*(t)=\Oh(c_{\max})$, in which a significant part of the moisture storage becomes saturated (if not, then a different nondimensionlisation should be considered). Let us introduce dimensionless quantities $c'(t)=\frac{c^*(t)}{c_{\max}}$ and $d'(t)=\frac{d(t)}{d_{\max}}$.

In order to balance temporal and precipitation terms in eqn~\eqref{eq:dim_c}, we rescale time as $t=\frac{c_{\max}}{P}t'$, obtaining a dimensionless version of this equation:
\begin{equation}
    \label{eq:dimless_c}
    \dtp{c'(t')}=1-\frac{d_{\max}}{P}d'(t')
\end{equation}
As long as $P$ is at least of the order $\Oh\left(d_{\max}\right)$ (which is a consequence of the earlier assumption that $c^*(t)=\Oh(c_{\max})$), then the last term does not have a dominant effect. Here, the characteristic timescale is therefore $t_c=\frac{c_{\max}}{P}$, which represents a characteristic timescale for the moisture storage dynamics.

In order to balance all terms in eqn~\eqref{eq:dim_sf}, we rescale time as $t=k_f t'$, and fast storage as $S_f=k_f\big(P-d(t)\big)S_f'$. After rescaling, the equation becomes:
\begin{equation}
    \label{eq:dimless_sf}
    \dtp{S_f'(t')}=1-\left(1-c'(t)\right)^b-S_f'(t')
\end{equation}
Now all terms are of order $\Oh(1)$. Like in the fully saturated case, the characteristic timescale of fast storage is $t_f=k_f$.

Finally, in order to balance all terms in eqn~\eqref{eq:dim_ss}, we rescale time as $t=k_s/d_{\max}t'$ and slow storage as $S_f=k_sS_s'$. The equation becomes:
\begin{equation}
    \label{eq:dimless_ss}
    \dtp{S_s'(t')}=d'(t)-\exp\left(S_s'(t')\right)
\end{equation}
All terms are of order $\Oh(1)$. Therefore, the characteristic timescale of slow store dynamics is $t_s=\frac{k_s}{d_{\max}}$.

\subsection{Fully saturated phase}

After reaching the time when the soil moisture storage reaches its maximum value $c^*(t)=c_{\max}$, the governing equations become:
\begin{subequations}
    \begin{align}
        \label{eq:after_saturation_sf}
        \dt{S_f(t)}&=P-d_{\max}-\frac{S_f(t)}{k_f} \\
        \label{eq:after_saturation_ss}
        \dt{S_s(t)}&=d_{\max}-\exp\left(\frac{S_s(t)}{k_s}\right)
    \end{align}
\end{subequations}
These two ODEs are independent of each other and can be solved analytically. The equation~\eqref{eq:after_saturation_sf} is identical to equation~\eqref{eq:saturated_case}, and likewise its solution is characterised by a single timescale $t_f=k_f$. The solution for~\eqref{eq:after_saturation_ss} is given by:
\begin{equation}
    S_s(t)=-k_s\ln\left(\frac{1}{d_{\max}}e^{-\frac{d_{\max}}{k_s}(t-C)}+\frac{1}{d_{\max}}\right)
\end{equation}
where $C$ is a constant that should match $S_s(t_s)$ from the solution for eqns \eqref{eq:dim_c}-\eqref{eq:dim_ss}. We can see that the characteristic timescale of this equation is $t_s=\frac{k_s}{d_{\max}}$. This can also be demonstrated by nondimensionalisation, with $\frac{k_s}{d_{\max}}$ as a scale of time and $k_s$ as a scale of groundwater storage $S_s$.

\section{Derivation of a short-time solution of the PDM}
\label{app:quasistatic_flow}

\noindent In this appendix, we derive a short-time solution for the surface flow in the PDM. We assume that the characteristic time for the fast store is much shorter than that for the soil moisture, \emph{i.e.} we consider the limit $\epsilon=\frac{t_f}{t_c}\rightarrow 0$.

Let us nondimensionalise equations~\eqref{eq:dim_c} and~\eqref{eq:dim_sf} by rescaling $t=t_ft'$, $d=d_{\max}d'$, $c=c_{\max}c'$, and $S_f=k_f\big(P-d(0)\big)S_f'$. Equation~\eqref{eq:dim_c} becomes
\begin{equation}
    \dtp{c'(t')}=\epsilon\left(1-\frac{d_{\max}}{P}d(t)\right)
\end{equation}
Note that in the limit $\epsilon\rightarrow 0$, this equation becomes $\dtp{c'(t')}=0$, \emph{i.e.} the soil moisture effectively does not change within the considered timescale. The solution is $c'(t')=c'(0)$, which is equivalent to $S'(t')=S'(0)$ and $d'(t')=d'(0)$.

After nondimensionalisation, equation~\eqref{eq:dim_c} becomes
\begin{equation}
    \dtp{S_f'(t')}=1-\left(1-c'(0)\right)^b-S_f'(t').
\end{equation}
The solution for this equation grows until reaching a steady-state value of
\begin{equation}
    \label{eq:sf_for_t_infinite}
    \lim_{t'\rightarrow\infty}S_f'(t')=1-\left(1-c'(0)\right)^b,
\end{equation}
or, after rescaling~\eqref{eq:sf_for_t_infinite} back to the dimensional variables,
\begin{equation}
    \lim_{t'\rightarrow\infty}S_f(t')=k_f\big(P-d(0)\big)\left[1-\left(1-\frac{c^*(0)}{c_{\max}}\right)^b\right].
\end{equation}
The corresponding surface water outflow is given by
\begin{equation}
    \label{eq:Qf_for_t_infinite}
    Q=\frac{S_f}{k_f}=\big(P-d(0)\big)\left[1-\left(1-\frac{c^*(0)}{c_{\max}}\right)^b\right].
\end{equation}
Note that, according to eqn~\eqref{eq:prob_distribution}, the term in the square bracket in~\eqref{eq:dq_pdm_app} is $F(c^*(0))$, which represents the fraction of the catchment's area that is fully saturated (see \cref{fig:PDM_surface_runoff_production}).

The initial condition for the soil moisture store, given by \eqref{eq:partially_saturated_ic}, is $S(0)=s_t+k_g P_0$. Therefore, using relation~\eqref{eq:S_from_C}, the flow~\eqref{eq:Qf_for_t_infinite} is expressed as
\begin{equation}
    \label{eq:early_time_Q}
    Q=\big(P-d(0)\big)\left[1-\left(1-\frac{s_t+k_g P_0}{S_{\max}}\right)^\frac{b}{b+1}\right].
\end{equation}
The difference between the initial $Q_0=P-d(0)$ and the flow given by~\eqref{eq:early_time_Q} is equal to
\begin{equation}
    \label{eq:dq_pdm_app}
    \Delta Q = Q-Q_0=(P-P_0) \left[1-\left(1-\frac{s_t+k_gP_0}{s_\text{max}}\right)^\frac{b}{b+1}\right].
\end{equation}
This formula has a simple physical interpretation. As we noticed before, the expression in the square bracket represents the fraction of the catchment that is fully saturated. Multiplying it by the difference in precipitation rate gives the fraction of rainfall that directly contributes to the surface water (fast store) without saturating the moisture storage first.

\end{document}